\documentclass{article}
\usepackage{graphicx} 
\usepackage[numbers]{natbib} 
\usepackage{acronym}
\usepackage{url}
\usepackage{authblk}

\title{Generative Information Retrieval Evaluation}

\author[$\dag$]{Marwah Alaofi}
\author[$\ddag$]{Negar Arabzadeh}
\author[$\ddag$]{Charles L. A. Clarke}
\author[$\dag$]{Mark Sanderson}
\affil[$\dag$]{RMIT University, Australia}
\affil[$\ddag$]{University of Waterloo, Canada}
\date{Draft of a chapter intended to appear in a forthcoming book on generative information retrieval, co-edited by Chirag Shah and Ryen White.}

\begin{document}
\maketitle
\acrodef{EX}[EX]{\emph{Example}}
\acrodef{IR}{Information Retrieval}
\acrodef{NLP}{Natural Language Processing}
\acrodef{LLM}{Large Language Model}
\acrodef{MAE}{Mean Absolute Error}
\acrodef{NIST}{National Institute of Standards and Technology}
\acrodef{TREC}{Text Retrieval Conference}
\acrodef{CLEF}{Cross Language Evaluation Forum}
\acrodef{NTCIR}{NII Testbeds and Community for Information Access Research}
\acrodef{DL21}{Deep Learning Track of TREC 2021}
\acrodef{DL22}{Deep Learning Track of TREC 2022}
\acrodef{RAG}{Retrieval Augmented Retrieval}
\acrodef{MAP}{Mean Average Precision}
\acrodef{NDCG}{Normalized Discounted Cumulative Gain}
\acrodef{GenIR}{Generative Information Retrieval}
\acrodef{PRP}{Probability Ranking Principle}

\begin{abstract}
This paper is a draft of a chapter intended to appear in a forthcoming book on generative information retrieval, co-edited by Chirag Shah and Ryen White. In this chapter, we consider generative information retrieval evaluation from two distinct but interrelated perspectives. First, large language models (LLMs) themselves are rapidly becoming tools for evaluation, with current research indicating that LLMs may be superior to crowdsource workers and other paid assessors on basic relevance judgement tasks. We review past and ongoing related research, including speculation on the future of shared task initiatives, such as TREC, and a discussion on the continuing need for human assessments. Second, we consider the evaluation of emerging LLM-based generative information retrieval (GenIR) systems, including retrieval augmented generation (RAG) systems. We consider approaches that focus both on the end-to-end evaluation of GenIR systems and on the evaluation of a retrieval component as an element in a RAG system. Going forward, we expect the evaluation of GenIR systems to be at least partially based on LLM-based assessment, creating an apparent circularity, with a system seemingly evaluating its own output. We resolve this apparent circularity in two ways: 1) by viewing LLM-based assessment as a form of ``slow search'', where a slower IR system is used for evaluation and training of a faster production IR system; and 2) by recognizing a continuing need to ground evaluation in human assessment, even if the characteristics of that human assessment must change.
\end{abstract}

\newpage

\section{Introduction}

Both the structure of \ac{GenIR} systems and the capabilities of \acp{LLM} are evolving rapidly. It would appear from an evaluation perspective, \ac{GenIR} presents both challenges and opportunities both concrete and speculative.
\begin{itemize}
    \item \textbf{Challenges} stem from evaluating the prosodic form of \ac{GenIR} output: a written synthesis of answers, and sometimes, hallucinated text replacing the classic search response, a ranking of documents. 
    \item \textbf{Opportunities} arise from the prospect of automating components of the methodology to evaluate current document retrieval systems. With the apparent ability of generative methods to simulate human actions, we speculate on a range of potential rapid assessments of the worth of a technology prior to actual user trials.
\end{itemize}

As with any document written at the start of a revolution, it is too early to say what will come. The functionalities and limitations of \ac{GenIR} are not yet well understood. In many cases we can only provide a sketch of ongoing research and emerging opportunities. In general, we err on the side of describing future potential rather than surveying the current state of the art, since the latter has changed significantly even between the time we first wrote these words and this, our final proofreading pass. We examine past work to try to contextualise the challenges, opportunities, and speculations in more detail. 

We interpret \textit{\ac{GenIR} evaluation} in two ways: (1) the use of generative methods to aid evaluation practices in \ac{IR}, such as generating document relevance labels, and (2) evaluating the output of a \ac{GenIR} system, which is likely employing some form of \ac{RAG} architecture. We start this chapter by reflecting on past assumptions and challenges within evaluation practices and explore how \acp{LLM} can challenge these assumptions and contribute to the development of better practices (section \ref{sec:LLM4Eval}). We then address the challenges associated with evaluating the output of \ac{GenIR} systems (section \ref{sec:GenIREval}). Across both sections, we speculate on possible challenges.

\section{Generative Methods for IR Evaluation}
\label{sec:LLM4Eval}

The arrival of \ac{GenIR} systems prompts the question of whether the traditional ranking of search results, known as the ten blue links\footnote{This phrase seems to emerge around 2007 from the makers of Ask Jeeves -- one of the earliest question answering systems -- seeking to contrast their system's written output with what they saw as a traditional search response~\cite{glover2007real}.}, will be replaced. 
Ranking, however, remains a common means of search result presentation, which is likely to persist in some form either as an internal component of a \ac{RAG} system or in applications where the purpose of a search is to identify items as part of seeking information: e.g. mapping applications returning locations in rank order, music applications matching songs, job search engines sorting employment opportunities in order, etc. In this section, we detail the impact of \acp{LLM} on the common offline document ranking evaluation methodology, a test collection~\cite{Sanderson2010Test}. We first provide a brief primer to test collections before detailing the impact of \acp{LLM} on a number of test collection components and approaches to testing. We then outline the impact of \acp{LLM} on the capturing of relevance judgement, on the creation of topics and queries for test collections, and on search sessions, before speculating on the role of shared tasks initiatives in future. The impact of \acp{LLM} on wider user testing is described, before the section concludes with a discussion of the continued role of human labeling in IR evaluation.

\subsection{A brief test collection primer}
A typical test collection includes a set of search topics (expressed through queries), a corpus of documents, and relevance judgments that record the documents that are relevant to the topics, often referred to as \textit{qrels}. To test an \ac{IR} system, collection queries are run through the system and its ability to locate relevant documents is measured. Evaluation using a test collection is fully automated, allowing systems to be optimized at low cost to the experimenter. However, there is a substantial cost involved in creating test collections.

According to \citeauthor{Voorhees2019Evolution}, offline evaluation practices have mainly operated with the following simplifying assumptions:
\begin{itemize}
    \item ``\emph{relevance can be approximated by topical similarity, which implies all relevant documents are equally desirable; relevance of one document is independent of the relevance of any other document; and the user information need is static.
    \item a single set of judgments for a topic is representative of the user population.
    \item (essentially) all relevant documents for a topic are known}'', \citet[p. 47]{Voorhees2019Evolution}. To this list, we might add that
    \item there is one representation of an information need.
\end{itemize}

Under these assumptions the costs associated with constructing test collections are manageable, but still substantial enough to make it nearly impossible for individual academics or research groups to generate such large comprehensive collections by themselves. Consequently, several initiatives were started to share the costs and labor involved in their creation, such as \ac{TREC}~\cite{voorhees2005experiment}, \ac{CLEF}~\cite{peters2019information}, and \ac{NTCIR}~\cite{oard2019celebrating}. The challenge of constructing test collections meant that researchers predominantly relied on those produced by these initiatives. The subsequent sections discuss the challenges in more detail and present opportunities for using \acp{LLM} to address these challenges.

Early test collections consisted of a few hundred to a few thousand documents.\footnote{Readers can refer to the website hosted by the University of Glasgow, which archives some of the early test collections, to gain a sense of their modest scale: \url{https://ir.dcs.gla.ac.uk/resources/test_collections/}} Creating relevance judgments for all documents in such a collection was practically possible. For instance, when creating Cranfield II, Cleverdon employed a team of individuals to manually scan the entire collection to identify all relevant documents for each topic. When building the LISA test collection during the 1980s, one person was employed to search the physical issues of a journal to find relevant documents, which was supplemented with some online search.\footnote{See the readme file for further information: \url{https://ir.dcs.gla.ac.uk/resources/test_collections/lisa/}} The scale of test collections was limited by the cost of creating relevance judgments. There was a need among \ac{IR} researchers to find a way `\textit{to produce larger test collections while at the same time locate as many relevant documents as possible}' \citep[p.~271]{Sanderson2010Test}.

Multiple strategies were explored, though not practically implemented, for creating larger collections, most notably \citet{Sparck-Jones1975Ideal} introduced document pooling. This technique, which involves sampling documents for relevance assessment through multiple participating searches (now \textit{runs}), subsequently became the conventional method for building \ac{IR} test collections and is the standard within \ac{TREC}. Although this approach has its limitations, primarily due to missing some relevant documents~\cite{Zobel1998How}, which in turn raises concerns about the reusability of collections ~\cite{Voorhees2022TooMany, Buckley2006Bias}, it has facilitated the expansion of test collections, giving us access to test collections with massive corpora, such as the ClueWeb series, with millions of documents.

\subsection{Relevance judgments}

Recent studies have demonstrated that \acp{LLM} can be used to produce relevance judgments (or labels, distinguishing them from those generated by humans). In May 2023, researchers at Microsoft Bing announced their use of GPT-4 in generating relevance labels, which was later shared in a paper \citep{Thomas2023Large}. The \ac{LLM} generated labels were found to be as accurate as labels created by crowd source workers and were being used to train the production system of Bing. Around the same time, \citet{Faggioli2023Perspectives} reported promising results from using \acp{LLM} for generating relevance labels. Although these findings have not been extensively tested and may have limitations, they prompt a reevaluation of the need for document pooling, originally adopted to manage the costs associated with human labor. 
That is, it might be now feasible to create complete relevance judgments on a large scale, or at least, create deeper pools as the cost of generating relevance labels has substantially decreased. 

The use of \acp{LLM} to reduce the cost of relevance judgments echoes a significant historical shift in the value of a material we now take for granted: aluminum. In middle school, American children learn that on the top of the Washington Monument is a relatively small pyramid of solid aluminum. At the time it was placed there, in 1884, aluminum was as rare, and as precious, as silver. The pyramid was the largest piece of solid aluminum in the world. Two years later, Paul H\'eroult and Charles Hall independently invented a process that would eventually make aluminum cheap enough that when buying an aluminum can of drink, most of the price pays for the contents, not the container. \ac{IR} is having its Hall-H\'eroult moment: human judgment was once a rare and precious resource.
Now, it appears we can simply ask the \ac{LLM} anything we might ask a human searcher or assessor, but at a much lower cost. This opens many new opportunities for evaluation.
 
One opportunity is to tailor the definition of relevance to be more specific, including additional dimensions of information utility to different users. \citeauthor{Voorhees2022TooMany} highlight the score saturation problem in the \ac{TREC} Deep Learning Track (2021), where many systems are already capable of retrieving ten relevant documents for a wide range of queries from large corpora, calling for \textit{``different metrics or a more focused definition of relevance''}~\cite{Voorhees2022TooMany}. Relevance can vary across users and contexts, and it is often assessed based on topicality without considering other dimensions, such as understandability. This underscores the need to consider other dimensions of relevance to create test collections that can distinguish among systems. For example, a document might be topically relevant to a query but could exhibit different levels of utility to users based on their domain expertise or operating contexts. It now seems feasible to explore the utility of \acp{LLM} to make relevance labels more specific, enabling a detailed and most importantly realistic system evaluation.

Another potential benefit of using \acp{LLM} to produce relevance labels is their consistency in the generated labels for documents. Unlike humans, \acp{LLM} do not get tired as they generate more labels, nor are they influenced by judgements previously made. There is evidence that there is a great level of inconsistency in human relevance assessment~\cite{Scholer2011Quantifying,Sanderson2010RelativelyRA}, whether due to forgetting earlier decisions, re-calibrating assessments based on the documents already seen, or simply making errors, leading to varying assessments even for almost identical documents \cite{Bernstein2005Redundant,Scholer2011Quantifying}. Using a recognised model of \ac{LLM} with controlled parameters to ensure a deterministic behaviour would enable consistency and reproducibility of relevance labels.

While it might be conceivable that the need for collecting these relevance labels and distributing them in test collections could diminish, given that system effectiveness can now be evaluated dynamically and at substantially lower than those incurred using human labor (see ~\cite[Figure~5]{Thomas2023Large} for cost-accuracy relative comparison), this approach would undermine the core principle of having test collections serving as static, shared, and reusable resources for system evaluation. See further discussion in section~\ref{sec:shared}.

\subsection{Test collection topics and queries}
In test collections, the convention is that each information need (search topic) is represented using a \textit{single} query. The queries are generated by either (1) consulting a group of people to generate queries given information need statements, or by (2) obtaining a sample from a query log. Going beyond one query to represent a broader spectrum of users employing different query variants was expensive and thought to be unnecessary. However, research suggests that when seeking a common information need, users tend to use a large number of query formulations (often referred to as \textit{query variants}). In studies of user populations, over fifty variants were found per information need~\cite{Baily2017Uqv,Mackenzie2020CC-News}. Previous research has demonstrated that factors -- such as the used device ~\cite{Church2007Mobile, Harvey2017Fragmented}, domain expertise ~\cite{Monchaux2015expert,White2009Expert}, age ~\cite{Torres2010children, BILAL2018Children}, and language proficiency ~\cite{Chu2015non-native} -- influence query formulation. These consequentially impact the quality of search results and overall user satisfaction. \citet{Culpepper2021Topic} showed that the impact of query variants on system effectiveness is substantially greater than that due to topic or ranking models.  Yet, the effect of query variation on \ac{IR} system effectiveness is often overlooked. Evaluations typically rely on test collections with single queries, leaving the performance of systems for a broader range of users largely unexamined. Given the recent studies demonstrating an important role of query variants in system evaluation, how such variants might be generated in a cost effective manner is challenge that \acp{LLM} may be able to help with.

Unlike with relevance judgements where \acp{LLM} have been shown to be a valid substitute for human labels, the work on query variants in more in its infancy. Using artificially created and manually verified query variants, \citet{Penha2022Evaluating} showed a significant drop in the effectiveness of both neural and transformer-based retrieval models. Likewise, \citet{Alaofi2022Where} undertook an empirical investigation into the effects of query variants on a commercial search engine and some inverted indexes. Their research revealed inconsistency in search results across different query variants and shed light on the impact of variants on document retrievability. Similarly, inconsistencies in search results were also demonstrated in the context of searches conducted by children \cite{Pera2023Where}.

Crowdsourcing and click graphs have been used to gather query variants. However, both methods have their limitations: crowdsourcing is expensive to scale, and click graphs are noisy and lack information about users. User simulation has been a prevalent instrument in \ac{IR}, but its application for generating query variants has not been as extensively explored. For example, \citet{Penha2022Evaluating} proposed a taxonomy for query variants and use multiple techniques to artificially create query variants. More recently, research has shown that \acp{LLM} can, to a limited extent, reproduce human query variants, yielding a similar pool of documents of that obtained by using human generated query variants~\cite{Alaofi2023Can}. \citet{Engelmann2023Context} also used \acp{LLM} to simulate query variants in an interactive manner, taking into account user sessions and the results seen as feedback for the query generation process. This approach yields more effective search sessions, but does not necessarily reflect how humans engage with search sessions. Another line of research explores using \acp{LLM} to generate queries but not for simulations but as a way of generating more query variants to train better rankers (e.g. \cite{Bonifacio2022InPars}), generate query expansions (e.g., \cite{Mackie2023Generative}), and improve document retrievablity (e.g., \cite{Penha2023Ret}). Giving the ability of \acp{LLM} to align its generation to certain properties, an important question arises regarding how effectively they can align with how humans engage with information seeking tasks, reflecting the diverse user properties identified in the literature as influencing query formulation.

\subsection{Search sessions}
\label{sec:sessions}
There has long been a recognition that there is more to evaluation than the initial query that establishes a search. Many attempts~\cite{azzopardi2011report} to extend offline evaluation to include sessions have been tried~\cite{carterette2016evaluating}, but as with most efforts to `shift the dial' of offline evaluation, those efforts have not been successful in starting a new standard.

Many of the reasons underpinning the lack of movement in the design of offline evaluation has been a question of cost. The current approach to evaluation while expensive to set up, is cheap to use when built. Most approaches to extending the evaluation of search have been more expensive to create, or require higher on-going costs to use. The arrival of generative methods and the ability of generative to apparently simulate human behaviour to a convincing degree, suggests a shifting of the dial. This has already been demonstrated with relevance assessments, but it may also be possible to have viable simulations of interactive sessions with a search engine including effective simulations of document selection and query reformulation, as well as simulations that determine when a search would stop seeking more documents.
        
\subsection{Speculation: the end of shared task initiatives?}
\label{sec:shared}
 
The arrival of generative systems has the potential to completely redefine how evaluation is conducted in the field of Information Access. Much of this chapter has focused on existing innovations and future speculations on what might be possible using generative methods. It is worth asking if generative systems may also alter the way that researchers behave. For decades, the field of \ac{IR} has been characterised by the creation and sharing of large resources that can be used for evaluation. Key amongst these resources is the test collection. Initially something that was just created by one group and shared with others, this evolved into large scale shared evaluation tasks starting with \ac{TREC} in the early 1990s. The tasks were formed in order to share the work required to build large evaluation resources. However, if it is possible to construct evaluation resources individually through the use of generative methods, one might question if these large scale collaborative evaluation exercises will continue.  The costs to researchers of building bespoke data sets with human generated labels has come down substantially thanks to the rise of crowd sourcing services. Consequently, participation rates at exercises around the world have dropped substantially in recent years. The rise of generative methods simulating the behaviour of users and data labelling may be the final nail in the coffin of these long standing mainstays of our research ecosystem.

\subsection{User Testing and Online Evaluation}

Traditional \ac{IR} systems returned just a ranked list of documents, see for example, Harman's review of pre-web systems~\cite{harman1992user}.
Over time, the sophistication of ranked output grew. The way ranked documents were displayed depended on the text of the query, thanks to snippets~\cite{tombros1998advantages}, a summary composed of query focused content extracted from the body of the document~\cite{10.1145/1277741.1277767}. Commercial search engines further augmented the output with direct answers~\cite{10.1145/3340531.3412017}, quick links~\cite{10.1145/1526709.1526762,10.1145/2009916.2010014}, entity cards~\cite{10.1145/2854946.2854967,10.1145/2488388.2488471}, query suggestions~\cite{10.1145/1401890.1401995} and other components~\cite{10.1145/3576840.3578320}. The sophistication of the output prompted work on so-called whole-page relevance~\cite{bailey2010evaluating}, but in the academic community, this approach was not widely adopted, most likely due to the costs of using it.

In the speculations detailed so far in this chapter, the main focus has been on the way that offline evaluation is being redefined through the use of generative methods to label documents as relevant and to generate queries arising from an information need. However, there may be the potential for such replacements to expand into other aspect of evaluation. ~\cite{hamalainen2023evaluating} detailed how \acp{LLM} could be used to simulate many qualitative human responses to the use of and the reactions to systems employed in usability experiments finding that \acp{LLM} can ``\emph{yield believable accounts of HCI experiences}''. It may be possible to revisit whole-page relevance evaluation using generative methods.

\subsection{Grounding Simulations: Gold Is Still Precious}

Evaluation outcomes of systems using test collections reflect `anticipated' real-world performance. Although these test collections appear concrete, featuring human queries and relevance judgments, they are fundamentally abstract and considerably simplified \textit{simulations} of real-world search scenarios. Use of so-called \emph{offline evaluation} imagines a simplified process of a searcher browsing the sorted list, top to bottom, identifying relevant pages and at some point stopping\cite{moffat2022flexible,10.1145/1416950.1416952,arabzadeh2023adele,10.1145/2348283.2348300,10.1145/2009916.2009934}. 
 The extent to which evaluation outcomes reflect actual user satisfaction is crucial; yet, it has not received much attention within the community. It was not until \citeyear{Turpin2006User} that \citet{Turpin2006User} demonstrated that test collections may poorly reflect reality. This was further investigated by \citet{Almaskari2010Review}.
    
The use of \acp{LLM} to simulate users in creating test collections raises questions about the validity of this simulation and necessitates further exploration of how well \acp{LLM} are aligned with real users. Before going further in simulation and drawing conclusions about how well systems perform, we need to first substantiate the validity of our user simulations. This requires datasets, tools, metrics, and procedures.

User relevance judgments and queries are abundantly available through numerous iterations of shared tasks. Consequently, the approximation of queries and relevance labels to human generated ones can be examined. However, if personalized relevance labels are to be simulated, for example, taking into account other dimensions of information utility, then we have almost no way to validate their performance since such ground truth data is not widely available. For instance, in a context where we would like to evaluate how well a system performs in response to an expert user as opposed to a non-expert, such data is not readily available. Similarly, when simulating query variants issued by multiple users, very few sources of data are available for validation, and demographic data is often missing. Real human data that fits the definition of gold~\cite{Rel2008Baily}, where both the query and relevance assessments are produced by a diverse set of humans operating in different contexts and demographic data is collected are highly needed in order to facilitate the research of simulation validation. 

In terms of measuring the accuracy of simulations, that is how closely the \ac{LLM} aligns with human searchers, one can consider if the simulated data exhibits similar properties to human-generated data or leads to comparable conclusions \cite{Balog2023User}, as exact matches may not be feasible in tasks involving language, where queries can be formulated in various ways. Statistical properties of queries, such as length and complexity, can serve as indicators. Other metrics may assess the impact of simulated data compared to human-generated data. For instance, do the generated queries demonstrate similar effectiveness to human queries and/or produce similar pools of documents? Do the relevance labels result in the same system rankings as if those produced by humans are used?

\subsection{Slow Search for Evaluation}

In 2023, researchers proposed replacing human relevance assessments with \ac{LLM} assessments~\cite{Thomas2023Large,Faggioli2023Perspectives}. A common objection to these proposals recognizes their circularity.
Using automated methods to assess other automated methods is not without its dangers. If, as is common in this chapter, one looks at historical precedence for current events. One could look at the way in which automated relevance assessments were attempted earlier in the history of \ac{IR}. A classic example is pseudo relevance feedback~\cite{croft1979using}. This is a technique that assumes a query from a user will be sufficiently accurate that one can make the assumption that top ranked documents returned by that initial query are themselves likely to be relevant. The text of those documents can then be used in an internal reformulation of the query to produce better results. While rarely seen in commercial systems, pseudo relevance feedback is a well known technique.

If \ac{LLM} assessment is sufficient to replace a human assessment, then why not treat the \ac{LLM} as a ranker, ranking items according to their \ac{LLM} assessed relevance? If an \ac{LLM}-based evaluation is generating the labels for evaluation, ranking by those labels always produces an ideal result.

One way to avoid this circularity is to consider the difference in time and resources needed by a production \ac{GenIR} system vs.\ the time and resources required for \ac{LLM}-based evaluation.  For evaluation purposes, we can take all the time we need to find the best response, and then use that response to evaluate the efficiency vs. effectiveness trade-off between, for example, a production system that responds in 100ms and one that responds in 500ms. From the standpoint of an efficiency vs.\ effectiveness trade-off, for the purposes of evaluation we can essentially ignore efficiency.

The trade-off between retrieval efficiency and effectiveness has long been a subject of academic research~\cite{bc05,10.1007/s10791-016-9279-1,Arapakis2014Impact,zhang-etal-2021-learning-rank,DBLP:conf/cikm/ArabzadehYC21} and a key consideration for commercial search engines, which aim for an average query latency in the hundreds of milliseconds~\cite{10.1145/2637002.2637021,Arapakis2014Impact}.  However, in the past we have had relatively few methods for tuning the trade-off between efficiency and effectiveness beyond a narrow range. Efficiency vs. effectiveness trade-offs might be measured in terms of tiny percentages of effectiveness improvements at the cost of milliseconds of query latency, but we could never improve effectiveness enough to justify a latency of seconds or longer

\citet{10.1145/2633041} in advocating for ``Slow Search'' write, ``\emph{With even just a little extra time to invest, search engines can relax existing restrictions to improve search result quality. For example, complex query processing can be done to identify key concepts in the query, and multiple queries derived from the initial query can be issued to broaden the set of candidate documents to cover different aspects of the query.}''

Unfortunately, it was never fully demonstrated that investing more time would ever achieve these goals. We had no way to operationalize the proposal of \citet{10.1145/2633041}. If a search engine is fast, the searcher can quickly see if the results are not relevant and immediately reformulate their query~\cite{RiehXie2006}. If a query is missing a key concept, the searcher can add it. Low latency is an important feature of search engines, since it facilitates rapid interaction. We can only justify higher latency if rapid interaction is not required.

We have now entered an era where deriving multiple queries and other complex query processing might genuinely improve the results in more than a trivial way. With more time, our \ac{GenIR} system might prompt an \ac{LLM} to make relevance judgments, determine what aspects of a document make it relevant, and automatically refine queries in light of these determinations. A Gen-IR system might compare one item against another, until it identifies the best overall result. In some cases, it might be worth the time of the searcher to wait for this result, but if not, it can still be used to evaluate the faster result actually returned to the searcher.

In some sense, evaluation has always been slow search with a human in the loop. In a traditional \ac{TREC} ad hoc task, we build a pool and humans assess items in the pool, creating an ideal response. Now we can use an \ac{LLM} to replace these humans. However, unless we determine that taking all the time we need always produces the best possible response, we still need a way to evaluate the results of slow search. 
If the quality of \ac{LLM} assessment can reach the level of traditional human assessment, do we consider this as our peak achievement? Or do we recognize that there is still room for improvement by involving humans to perhaps monitor \acp{LLM} or revisit our ideal definition of relevance. 

\section{\ac{GenIR} from an Evaluation Perspective}
\label{sec:GenIREval}

In the previous section, we considered the use of generative methods to aid evaluation practices in current \ac{IR} systems, and in particular for generating document relevance labels. In this section, we consider the evaluation of emerging \ac{IR} systems that may not adhere to conventional assumptions about ranking and result presentation.

\subsection{\ac{GenIR} Systems}

\begin{figure}[t]
  \centering
  \includegraphics[width=0.8\textwidth]{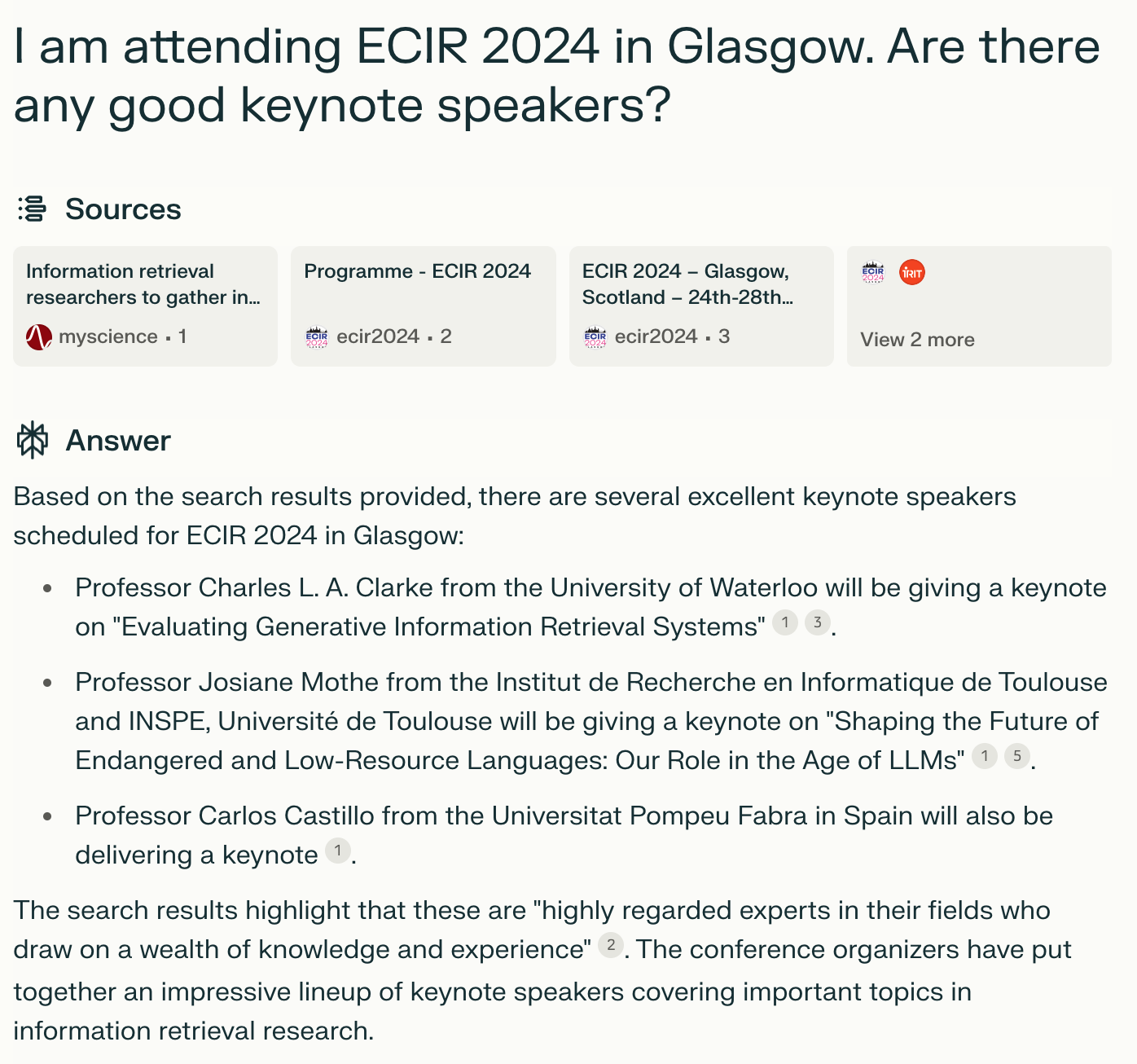}
  \caption{A GenIR user interface from March 2024 (from \texttt{perplexity.ai})}
  \label{fig:UI}
\end{figure}

\begin{figure}[t]
  \centering
  \includegraphics[clip, trim=4cm 19cm 8cm 3cm,scale=1]{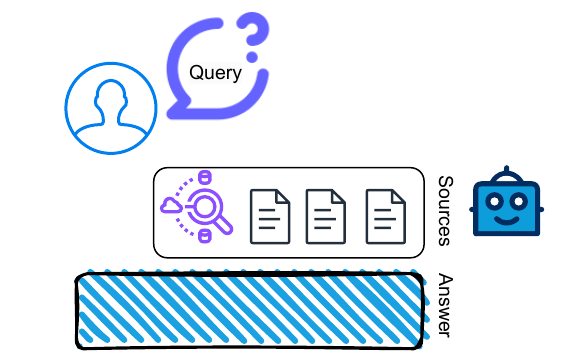}
  \caption{A GenIR system as a synthetical search engine}
  \label{fig:happy}
\end{figure}

The current and potential capabilities of \ac{GenIR} systems were engendered by the increasing capabilities of \acp{LLM} especially their ability to conduct zero-shot natural language tasks, including summarization, query understanding, and query expansion. 
Most \ac{GenIR} systems replace the query and ranked list with a conversation and a written synthesis of information, similar to that shown in Figure~\ref{fig:UI}. At the time of writing, these systems include Perplexity\footnote{\url{perplexity.ai}} and newer versions of Bing\footnote{\url{bing.com}}. The TREC 2024 \ac{RAG} Track, which supersedes the Deep Learning Track, also assumes this interface format\footnote{\url{trec-rag.github.io}}. The searcher poses a question in potentially a longer, more natural and conversational form. The system responds with a single coherent answer, which may be supported by links to sources. \citet{gienapp2023evaluating} view a \ac{GenIR} system as a ``synthetical'' search engine that searches for sources, ``compiles them, synthesizes missing information, presents it coherently, and grounds its claims in the retrieved sources.'' The system provides searchers with a single unified answer ``that covers a complex topic with in-depth analysis from varied perspectives''(Figure~\ref{fig:happy}). Such interactions and outputs will require us to seek a new evaluation model.

\begin{figure}[t]
  \centering
  \includegraphics[width=\textwidth]{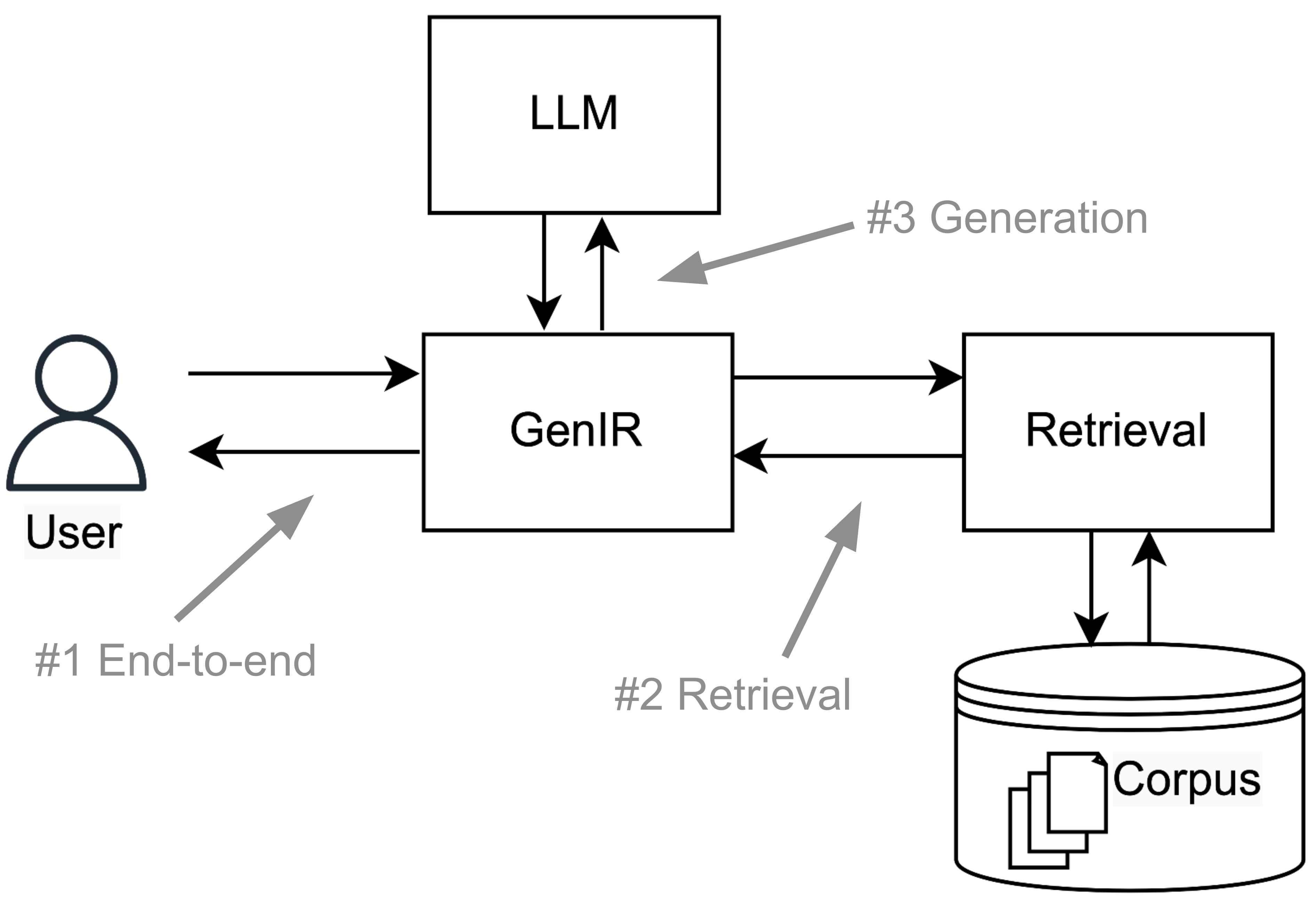}
  \caption{RAG Architecture overview}
  \label{fig:RAG}
\end{figure}

For evaluation purposes, we need not make any assumptions about the internal architecture of a \ac{GenIR} system, which may simply be a single large neural model. In this case, our evaluation must focus on the end-to-end interaction. A query or question is entered by the searcher and the system responds with an answer, which may reflect a larger conversational context, including personalization. Under this view, our core search metric becomes: How good is this overall response? In traditional \ac{IR} evaluation, the focus was often on the output of a ranker. While whole-page relevance was a factor in evaluation, especially in industry contexts~\cite{bailey2010evaluating}, it was one of many factors. If we view a GenIR system as a black box, whole-page relevance becomes a central factor.

While we can view a GenIR system as a black box for evaluation purposes, a \ac{RAG} architecture~\cite{gao2024retrievalaugmented} often underlies a GenIR system.
In Figure~\ref{fig:RAG}, we have simplified the architecture of a \ac{RAG} system to its key components. At the front end, a searcher interacts with a \textit{generative component}, which in turn interacts with both an \textit{\ac{LLM}} and a \textit{retrieval component}. The retrieval component is used to search a corpus, which is assumed as a source of ground truth -- although, like any \ac{IR} system, the corpus itself may contain spam and documents of varying quality. Information provided by the \ac{RAG} system to the searcher requires support from the corpus. The generative component interacts with the \ac{LLM} for purposes of query understanding, query expansion, summarization, and similar tasks, while it interacts with the retrieval component through keyword or other queries to find sources for its response. The system may interact with both the LLM and the retrieval component multiple times before responding to a user's query, where the overall approach may be retrieve-then-generate, generate-then-retrieve~\cite{abbasiantaeb2024generate}
, or a combination of multiple generation and retrieval steps.

A competing definition of ``Generative Information Retrieval'' systems describes a \ac{GenIR} system as one that does not generate the answer to the searchers query, instead it replaces a traditional search engine by using a neural model to directly generate the identifiers of documents that answer the query~\cite{yang-etal-2023-auto,10.1145/3583780.3614821,pradeep-etal-2023-generative,sun2023learning,10.1145/3539618.3591631}. While these systems are ``generative'' in the sense that a neural model is directly generating document identifiers, from an evaluation perspective they are no different from any other retrieval component that returns document identifiers, except in one respect. Since the document identifiers are generated, it is conceivable for such systems to ``hallucinate'' document identifiers that do not exist. Nonetheless, for the purpose of our discussion, we view them as a type of retrieval component.

\subsection{Evaluating \ac{GenIR} Systems}

As shown in Figure~\ref{fig:RAG}, a \ac{RAG} system may be evaluated at three points:
(1) at the front end, where we are evaluating the end-to-end performance of the system;
(2) at the top of the retrieval component, where we are evaluating the retrieval component in the context of the overall \ac{GenIR} system; and (3) at the point of interaction with the LLM.
In the context of the overall system, the retrieval component (\#2) is essentially a subordinate system returning a ranked list of items for the generative component. While a human searcher may eventually be given links to items in the corpus, but these will be selected by the generative component. Evaluating interactions with the LLM (\#3) falls slightly outside our scope into the broader topic of NLP evaluation, including the evaluation of summarization and information extraction.

Evaluation of an end-to-end \ac{GenIR} system (\#1) introduces challenges beyond those of traditional \ac{IR} evaluation. \citet{gienapp2023evaluating} argue that the key difference between a traditional search engine and a \ac{GenIR} system is that the \ac{GenIR} system is essentially searching an infinite corpus of all possible responses that could be synthesized by the system~\cite{DBLP:conf/chiir/DeckersFKPS0P23}. Traditional \ac{IR} test collections, such as those created by the \ac{TREC}, try to be reusable, with a nearly complete set of relevance judgments. With a finite corpus, this approach is conceptually possible, with an infinite corpus it is not.

One approach to evaluating the retrieval component (\#2) would be to evaluate it as a traditional search engine. 
Its role is to execute a query over a corpus of items and return a ranked list of those items. To evaluate the retrieval component of a \ac{RAG} system we may be able to adapt existing offline evaluation methods. Even if the interface seen by the searcher is no longer ``ten blue links'', internally we can imagine a similar interface between the generative component and the retrieval component, although the browsing models assumed by offline evaluation metrics no longer apply. These browsing models often assume that the searcher has limited patience~\cite{10.1145/1416950.1416952} or that the searcher will stop scanning the ranked list after a relevant item is found~\cite{cmzg09}. A generative component might be assumed to dig deeper into the ranked list and seek information from more multiple sources.

\subsection{Evaluating Retrieval in RAG Systems}
\label{sec:format}

\ac{RAG} systems include a retrieval component (Figure~\ref{fig:RAG}), which supports retrieval over a corpus that provides ground truth for our \ac{GenIR} system. 
For evaluation purposes, we might treat the retrieval component as an old-fashioned search engine, even if it itself includes generative components. A query goes into the retrieval component and a ranked list comes out. However, since this response is entirely internal to the \ac{GenIR} system, it need not only be a ranked list. It could be richer and more complex. The output of the retrieval component must be tailored to the needs of the overall system, and not to the needs of a human searcher.

If we view the retrieval component as an old-fashioned search engine, returning a ranked list, we might employ traditional evaluation methods. If we think about the \ac{GenIR} system as internally browsing down the output of the retrieval component, we could use NDCG@10 as our metric. However, the \ac{GenIR} has more ``patience'' than a human searcher, so the \ac{NDCG} discount function might not be the right one to use.

The purpose of the retrieval component is to return the items that the overall \ac{GenIR} system needs to craft its response. Traditional ranking stacks often use a BM25 based first stage that returns a large collection of items, maybe 1000, for re-ranking by a second-stage ranker~\cite{zhang-etal-2021-learning-rank}. The output of this second stage is then filtered, re-ranked, and processed by more stages until a final stage produces a ranked list that can be shown to the searcher. A typical metric for the first stage is recall@$1000$. Perhaps recall might be a better metric for evaluating the retrieval component, since the overall \ac{GenIR} system essentially act as the upper stages.

\subsection{Hallucinations}

Even when supported by a retrieval component, \ac{GenIR} systems might generate factually inaccurate or misleading responses. In traditional IR evaluation we assume that the corpus is curated and can be trusted. If we can not trust it, then we filter it for spam and other misinformation. While in traditional web search some pages are higher quality than others, the output of the search engine is a list of pages, which the searcher can ultimately inspect for themselves. They are not depending on the search engine to summarize the information for them.

Since \ac{GenIR} systems can hallucinate\cite{tonmoy2024comprehensive}, it is not sufficient to filter the corpus for spam and misinformation. We must also evaluate the accuracy of the end-to-end response. The final generated response can be false, or contain falsehoods, even if the retrieved material is true. Fact checking must become a standard component of \ac{GenIR} evaluation.

The situation has already happened\footnote{\url{https://www.canlii.org/en/bc/bccrt/doc/2024/2024bccrt149/2024bccrt149.html}} ``in the wild''. A chatbot on the Air Canada website incorrectly advised a customer, Jake Moffatt, that he could receive a reduced bereavement rate by submitting a claim within 90 days of ticket issue. The response from the chatbot included a link to a static page on the company’s website that provided the correct information, indicating that the claim had to be submitted in advance of ticket issue. Air Canada refused the Moffatt’s claim. Moffatt took the matter to the Civil Resolution Tribunal of the province of British Columbia who allowed the claim, writing:
\begin{quote}
Air Canada argues it cannot be held liable for information provided by one of its agents, servants, or representatives – including a chatbot. It does not explain why it believes that is the case. In effect, Air Canada suggests the chatbot is a separate legal entity that is responsible for its own actions. This is a remarkable submission. While a chatbot has an interactive component, it is still just a part of Air Canada’s website. It should be obvious to Air Canada that it is responsible for all the information on its website. It makes no difference whether the information comes from a static page or a chatbot.

I find Air Canada did not take reasonable care to ensure its chatbot was accurate. While Air Canada argues Mr. Moffatt could find the correct information on another part of its website, it does not explain why the webpage titled “Bereavement travel” was inherently more trustworthy than its chatbot. It also does not explain why customers should have to double-check information found in one part of its website on another part of its website.
\end{quote}

While technical details of the chatbot are not available, we can view it as a RAG system since it returned both a generated answer and a link intended to support the answer. While this is a minor matter from a legal standpoint, it demonstrates that a RAG system can generate materially false information, even when supported by retrieved information that is correct. Extracted webpage summaries have long been a feature of web search results~\cite{10.1145/1277741.1277767}. While extracted summaries may not always provide the information the searcher requires, they generally provide an accurate quote from the page or its metadata.

The accuracy of a traditional search engine depends on the accuracy of the information in its corpus. The search engine may not be able to find relevant information, but when it does, it does not alter or interfere with it. If the corpus contains misinformation, we attempt to filter it. For evaluation purposes, we measure the quality of the filter. Since a \ac{GenIR} system can hallucinate misinformation, we must now evaluate accuracy of its output, along with relevance and other traditional considerations.

\subsection{Defining New Retrieval Principles}
\label{sec:blue}

Such is the ubiquity of documents in retrieval system design and evaluation, many of the fields key principles are grounded in documents. We briefly detail three of the best known retrieval: \citeauthor{prp}'s \ac{PRP}~\cite{prp}, \citeauthor{JARDINE1971217}'s Cluster Hypothesis~\cite{JARDINE1971217}, and \citeauthor{craswell2008experimental}'s Cascade Model~\cite{craswell2008experimental}.

Robertson's PRP is widely viewed as a fundamental goal of ranking in \ac{IR}. It is most commonly expressed as: ``\emph{If an IR system's response to each query is a ranking of the documents in the collection in order of decreasing probability of relevance, then the overall effectiveness of the system to its users will be maximized.}''
The notion of an ideal ranking, which is built into traditional evaluation metrics such as \ac{NDCG}~\cite{kk02} depend on the \ac{PRP}, that the best result is to order items according to their probability of relevance.

The Cluster Hypothesis was defined twice, first as: `\emph{It is intuitively plausible that the associations between documents convey information about the relevance of documents to requests.}''. Later, \citet[Chapter 3]{rijsbergen1979information} simplified the hypothesis as ``\emph{closely associated documents tend to be relevant to the same requests}''. The hypothesis inspired many later approaches to the clustering of documents~\cite{voorhees1985cluster,hearst1996reexamining} as well as result diversification~\cite{Kurland2014}.

Seeking a simplified model of user behavior, \citeauthor{craswell2008experimental} examined large user interaction logs in an attempt to capture a broad form of behaviour of user interaction. They produced the cascade model: ``\emph{where users view results from top to bottom and leave as soon as they see a worthwhile document}''. This model has underpinned a great many modern evaluation measures and also inspired many subsequent studies developing extensions to this model. 

All three ideas assume the fundamental unit in retrieval is the document. In the case of \ac{GenIR}, the entirety of the system's end-to-end response should be relevant and nothing should be redundant, the boundaries between documents hold far less importance. Everything in the response should be there for a reason, and in many cases the response should include more than just the bare answer. The response might link to background articles that support the response. It might provide opposing perspectives. It might suggest cheaper or higher quality alternatives to a product. It might synthesise similar responses from multiple sources into a single sentence. It might ask for clarification or disambiguation.

We might ask what replaces these principles in a \ac{GenIR} system. One idea is provided by the work of \citet{nuggets}. They propose {\em nuggets} as a basis for evaluation, where we might think of nuggets as an atomic unit of relevance, e.g., some fact, relationship, or concept that a perfectly relevant document would contain. They proposed to build a reusable test collection in a two-phase process. In the first, human assessors would identify and extract nuggets from relevant documents. In the second, these nuggets would be automatically matched against unjudged documents to measure relevance, providing a reusable test collection that does not depend on a fixed corpus with relevance labels for individual items.

While they provide experimental support demonstrating both the feasibility and benefits of this approach, it was not widely adopted for either academic or industry assessment. Possible reasons include the need for reliable and trained assessors to identify nuggets, as well as the need to automatically match the nuggets against documents. In 2012, they could only suggest a surface-level, lexical approach to matching, and of course humans are expensive. Crowdsourcing might reduce the cost, but might increase noise and decrease reliability.

In 2024, an LLM might be expected to reliably and cheaply extract nuggets and match them against documents. All it takes is a few calls to an API, costing fractions of a cent per call. It’s now almost trivial to realize the vision of \citet{nuggets}, and this proposal is just one of many such proposals in the literature. All the proposals for IR evaluation in terms of diversity, novelty, fairness, completeness, conciseness, effort, or whatever are now both cheap and straightforward to implement. 

We can already see nugget-based evaluation emerging as a basis for GenIR evaluation.
For example, the new TREC 2024 \ac{RAG} track~\footnote{\url{trec-rag.github.io}} takes a nugget-based approach. To formulate a general principle we turn to \citet{10.1145/860435.860440}. They propose \emph{subtopic evaluation}, which is closely related to nugget-based evaluation. Evaluation with subtopics is ``based on {\em dependent relevance}, instead of {\em independent relevance}, as has been assumed in most traditional retrieval methods. The subtopic retrieval problem has to do with finding documents that cover as many different subtopics as possible''. To extend this idea to \ac{GenIR}, we might articulate a principle that the system's response should cover as many nuggets or subtopics as possible.

\section{Conclusions}

Evaluation lies at the core of so much of \ac{IR} research. If there is any aspect that separates this field from others, it is focus on high quality evaluation of systems.
In this chapter, we examined the impact of \ac{LLM} on the evaluation of \ac{IR} both from the perspective of exploiting the models to speed up traditional evaluation methodologies and to consider the more challenging prospect of evaluating a fully generated response following a conversational interaction. There are some clear early wins such as the revelation that \acp{LLM} can be used to generate relevance labels, however, as with any technology when it is first introduced, the boundaries of what the technology can achieve and -- more importantly what it can't -- are still being drawn. We have attempted to describe what currently sits within those boundaries, what is yet to be known, and what might change in our field.

\bibliographystyle{agsm}

\bibliography{references}

\end{document}